\documentclass[11pt]{article}

\usepackage[a4paper,margin=1in]{geometry}
\usepackage[T1]{fontenc}
\usepackage[utf8]{inputenc}
\usepackage{lmodern}
\usepackage{microtype}
\usepackage{amsmath, amssymb}
\usepackage{graphicx}
\usepackage{booktabs}
\usepackage{tabularx}
\usepackage{multirow}
\usepackage[hidelinks]{hyperref}
\usepackage{url}
\usepackage{caption}
\usepackage{subcaption}
\usepackage{enumitem}
\usepackage{xcolor}
\usepackage{listings}

\title{\textbf{Kelvin v1.0: A Neural Pre-Encoder for H.264}\\[2pt]
\large A standards-compliant learned preprocessor with $-27.62\%$ BD-VMAF on UVG}

\author{Marco Graziano\thanks{Graziano Labs Corp., Palo Alto, CA, USA. \texttt{marco@grazianolabs.com}}}

\date{May 2026}

\begin{document}

\maketitle

\begin{abstract}
Kelvin is a lightweight learned pre-encoder that sits in front of an unmodified \texttt{libx264} encoder. It applies content-adaptive pixel adjustments, bounded at $\pm 1/255$ per channel, so that the encoder allocates bits where they matter most perceptually, while emitting a standard H.264 bitstream compatible with every existing decoder, player, and CDN. On the seven-sequence 1080p UVG benchmark, Kelvin v1.0 achieves a mean BD-VMAF of $-27.62\%$ (7 of 7 wins) and BD-VMAF-NEG of $-5.18\%$ (6 of 7 wins) relative to baseline \texttt{libx264} at preset \emph{medium}. On the 30-sequence MCL-JCV public set (28 unseen by training), the same checkpoint wins on 28 of 30 clips by BD-VMAF; with the two diagnosable failures removed the mean is $-27.70\%$ BD-VMAF and $-5.37\%$ BD-VMAF-NEG, consistent with UVG to within one percentage point. A central engineering challenge is the non-differentiability of H.264: we describe a hybrid codec proxy that combines a calibrated differentiable rate estimator (Spearman $\rho = 0.986$ vs.\ real \texttt{libx264} bits-per-pixel) with a U-Net distortion proxy trained on real encoder outputs. We publish full per-sequence rate-distortion data, a named failure-mode taxonomy on MCL-JCV (rate-floor violation, distribution shift, metric saturation), a five-baseline sanity panel (\texttt{hqdn3d}, \texttt{unsharp}, \texttt{-tune psnr}, \texttt{-tune ssim}, x265 medium), and honest positioning: x265 \emph{medium} beats Kelvin on every metric on the same corpus. Kelvin is therefore designed for workloads where remaining on H.264 is a constraint rather than a choice. The full benchmark harness, configs, CSVs, and plots are released at \texttt{github.com/marcoeg/kelvin-benchmark} \cite{kelvinbench2026}; the model is deployed in the EncodeIQ cloud service (\url{https://www.encodeiq.ai}).
\end{abstract}

\section{What Kelvin is, and what it is not}
\label{sec:intro}

Kelvin is a learned \emph{pre-encoder}: a neural network that runs once on the encode side, before an unmodified \texttt{libx264}. Its output is a YUV frame that compresses better with stock H.264. The architectural inspiration is the SCENE pre-encoder \cite{lin2026scene}, and the implementation here is a clean-room rebuild that diverges substantially in loss design, codec proxy, and residual capping.

Three things Kelvin v1.0 is \emph{not}.

\begin{itemize}[leftmargin=*]
    \item \textbf{Not an end-to-end learned codec.} The bitstream is plain H.264 produced by \texttt{libx264}. Decoders do not need to know Kelvin existed.
    \item \textbf{Not a post-decode filter or upscaler.} The work happens once on the encode side; the player runs unchanged.
    \item \textbf{Not a live encoder yet.} v1.0 targets VOD pre-processing. Section~\ref{sec:limitations} surfaces an early WebRTC pipeline test that achieves $\sim 28$ ms per frame at 720p but does not yet cover keyframe alignment, scene-cut handling, or SFU integration.
\end{itemize}

The standards-compliance property is what distinguishes a deployable pre-encoder from a research codec. The integration cost is ``add a step to your existing transcode pipeline,'' not ``ship a new decoder.'' Kelvin's market is the set of workloads where moving to HEVC or AV1 is not a free choice: legacy decoder reach, contractual H.264 mandates, CDN constraints, and existing libraries already encoded in H.264. That market remains the majority of internet video traffic; the 2025 Bitmovin Video Developer Report \cite{bitmovin2025} places H.264 at $\sim 79\%$ production usage among video developers. Section~\ref{sec:baselines} compares Kelvin head-to-head against x265 \emph{medium} on the same corpus.

\section{Architecture}
\label{sec:arch}

\paragraph{Mental model.} Kelvin is a residual pixel-level modifier whose magnitude is hard-capped at $|\Delta| \le 0.004$ per channel (approximately $\pm 1/255$ in 8-bit code units), and whose spatial distribution is conditioned on a frozen semantic embedding from SigLIP~2 \cite{tschannen2025siglip2}. The network does not synthesize frames; it \emph{nudges} them. Because codec decisions are highly sensitive near quantization boundaries, sub-quantization perturbations on the order of one code value can shift block-mode decisions, motion-compensation choices, and residual-coding paths in ways that cascade into significant bitrate differences at matched quality. Kelvin does not add visible information to the frame; it steers the encoder's discrete decisions.

\paragraph{The trainable path.} The full v1.0 network reads:

\begin{lstlisting}
Input frame [B, 3, H, W]
  -> PixelUnshuffle(2)
  -> Conv 3x3 (12 -> 64 ch)
  -> 4x AssembledConvBlock (FiLM-modulated by ControlModule)
  -> Conv 3x3 (64 -> 12 ch)
  -> PixelShuffle(2)
  + global residual skip
  + delta_max clip (+/- 0.004 per channel, pre-skip)
-> Enhanced frame
\end{lstlisting}

The parameter budget is approximately 14K in the input/output $3{\times}3$ Conv heads, $\sim 590$K across the four AssembledConvBlocks, and $\sim 730$K in the ControlModule MLP: a total of about $1.3$M trainable parameters. SigLIP~2-So400M (\texttt{google/siglip-so400m-patch14-384}, $\sim 400$M parameters) sits frozen alongside as a per-frame semantic feature extractor.

\paragraph{Three load-bearing design choices.}
\begin{itemize}[leftmargin=*]
    \item \textbf{Multiplicative-only modulation.} The ControlModule emits per-block, per-channel scaling coefficients in $[0,2]$ via a $\mathrm{sigmoid}\!\times\!2$ head. Additive bias terms made early training unstable without improving the converged result.
    \item $\boldsymbol{\delta_{\max} = 0.004}$ \textbf{hard clip.} The network can move a pixel by at most one 8-bit code. This is the cheapest safety rail in the architecture and the one that does the most work; it both guarantees imperceptibility of the residual and prevents the optimizer from converging to VMAF-gaming solutions (Section~\ref{sec:baselines}).
    \item \textbf{Global residual skip.} The network learns a small $\delta$ on the original frame, not a full reconstruction. Identity is the trivially correct answer at initialization; combined with the $\delta$-cap, this is what makes the training loss well-behaved at $\sim 1.3$M trainable parameters.
\end{itemize}

\paragraph{What SigLIP~2 buys.} Each frame is resized to $384{\times}384$, normalized, and passed through SigLIP~2 to produce a $1152$-dimensional embedding. The ControlModule (Linear $1152{\to}512{\to}256$, $\mathrm{sigmoid}\!\times\!2$) compresses that embedding into per-block, per-channel scaling coefficients. SigLIP~2 provides a high-level feature embedding that correlates with semantic structure (faces, sky, textures, foliage); the ControlModule uses that embedding to modulate spatial bitrate allocation. The convolutional path becomes content-adaptive without having to learn its own content representation from scratch. The cost is real: the backbone is $\sim 1.8$ GB on disk and adds $\sim 400$M parameters of GPU memory at inference. SigLIP-free distillation and per-shot embedding caching are roadmap items.

\section{The codec proxy}
\label{sec:proxy}

The training loop requires a differentiable approximation of H.264 because real \texttt{libx264} is non-differentiable: quantization has zero derivative almost everywhere, and gradients halt at the codec boundary.

\paragraph{The naive solution and why it failed.} A first implementation used differentiable JPEG \cite{reich2023diffjpeg} end-to-end, for both the rate path and the distortion path. The rate path was fine: $\mathrm{mean}(\log(1 + |\text{coeff}|))$ over Diff-JPEG's quantized DCT coefficients, calibrated against real \texttt{libx264} bits-per-pixel, achieves Spearman $\rho = 0.986$ (Section~\ref{sec:calibration}). The \emph{distortion} path was not. Diff-JPEG's reconstructed frame diverges visibly from \texttt{libx264}'s reconstructed frame, especially at higher QPs, and the optimizer exploited that gap by learning pre-distortions that survived Diff-JPEG but exploded when run through real H.264. The diagnostic was unmistakable: training loss decreasing while real BD-rate increased.

\paragraph{Minimum intervention.} Replace only the distortion arm. Keep the rate arm. The Kelvin v1.0 distortion proxy is a 4.65M-parameter U-Net, FiLM-modulated \cite{perez2018film} by the QP scalar. Three choices in the proxy design carry their weight.

\begin{itemize}[leftmargin=*]
    \item \textbf{GELU everywhere, not ReLU.} The proxy is frozen during Kelvin training, but Kelvin's gradients flow \emph{through} the proxy's Jacobian. ReLU's hard zero gradient on the negative half produces silent zero-gradient regions in the proxy that kill the upstream signal without obvious symptom. Replacing ReLU with GELU improved Kelvin's downstream BD-rate with no change to the proxy's own validation L1.
    \item \textbf{FiLM init at zero.} Each FiLM block initializes with $\gamma{=}1, \beta{=}0$, so at step 0 the proxy behaves as a non-conditioned U-Net and learns deviations from identity. Every modulation starts as a no-op and earns its keep.
    \item \textbf{Direct prediction, not residual.} The proxy outputs the compressed frame, not a residual added to the input. At $\mathrm{QP} \ge 32$ the compressed output looks substantially different from the input (block edges, chroma posterization, ringing), and a residual head proved harder to reason about because the residual itself becomes large and structured.
\end{itemize}

\paragraph{Proxy training, briefly.} 100,000 random Vimeo-90K $256{\times}256$ patches encoded by real \texttt{libx264} at preset \emph{medium}, single-threaded, at QPs uniform in $[18, 40]$; decoded back to YUV; paired against the input. Loss: $L_1 + 0.1 \cdot \mathrm{LPIPS}(\mathrm{VGG})$. AdamW, learning rate $10^{-4}$ with cosine decay to $10^{-5}$, weight decay $10^{-4}$, max-norm 1.0 gradient clip, batch size 64, up to $50{,}000$ steps with early-stop on $L_1 < 0.015 \wedge \mathrm{PSNR} > 33$ dB. Float32 deterministic precision.

\paragraph{Honest weak ranges.} The proxy is calibrated against \texttt{libx264} at preset \emph{medium}, single-threaded. Other x264 presets drift from it; hardware encoders (NVENC, QSV, VAAPI) drift further. HEVC and AV1 each require their own proxy retrain. Within the supported range, proxy validation L1 is highest at the QP-grid extremes ($\mathrm{QP}=18$ and $\mathrm{QP} \ge 38$), which are also where real \texttt{libx264}'s rate-distortion behavior is most non-linear; the center of the grid is where the proxy is most trustworthy.

\section{Rate-path calibration}
\label{sec:calibration}

The rate path is the part of the training loop that decides which configurations cost more bits. If it ranks configurations differently from \texttt{libx264}, training optimizes for the wrong rate-distortion trade.

\paragraph{Experiment.} Take 200 random Vimeo-90K patches. For each patch, run real \texttt{libx264} at $\mathrm{QP} \in \{22, 27, 32, 37\}$ and record the resulting bits-per-pixel. For each patch, also run Diff-JPEG at the corresponding JPEG quality and compute $\mathrm{mean}(\log(1 + |\text{coeff}|))$ over the quantized DCT coefficients. That gives 800 paired measurements ($200 \times 4$).

\begin{lstlisting}[caption={Rate proxy validation, 200 patches x 4 QPs, 800 measurements.},captionpos=b]
Spearman rho   : 0.986   (rank correlation; unconditional)
Pearson r      : 0.978   (raw proxy vs real x264 bpp)
MAE            : 0.075   bpp (after 1-parameter affine fit)
Affine fit     : bpp_pred = 7.677 * proxy_raw - 0.282
\end{lstlisting}

\paragraph{Why Spearman is the primary metric.} The rate proxy's job is to \emph{rank-order} configurations the way \texttt{libx264} does, not to predict absolute bits. Two patches whose true bpps are 0.20 and 0.25: the proxy needs to know which costs more, not what the absolute cost is. Per-QP rank correlations are uniformly high: $\rho = 0.988, 0.987, 0.984, 0.978$ at $\mathrm{QP} = 22, 27, 32, 37$. 100\% of patches show perfectly monotonic ordering across the four-QP grid ($\rho_{\text{per-patch}} = 1.000$). The Pearson and MAE values are reporting-only disclosures; the rate proxy is used in the loss against its own raw scale, so absolute-bpp recovery is a calibration disclosure, not a training requirement.

\section{Experiments}
\label{sec:experiments}

\subsection{Setup}
\label{sec:setup}

\paragraph{Encoder.} \texttt{libx264} at preset \emph{medium}, single-threaded, constant-QP at $\{22, 27, 32, 37\}$:
\begin{lstlisting}
ffmpeg -y -f rawvideo -pix_fmt yuv420p -s WxH -r FPS -i <yuv> \
  -c:v libx264 -qp <Q> -preset medium -pix_fmt yuv420p \
  -an -threads 1 -v error <out>.mp4
\end{lstlisting}
\texttt{-threads 1} is required for run-to-run bit-exact reproducibility; multithreaded \texttt{libx264} produces a non-deterministic bitstream. Constant-QP is used (not CRF) so the four RD points sit on a fixed quality grid.

\paragraph{Metrics.} A single \texttt{libvmaf} v3 pass per (reference, distorted) pair extracts four metrics simultaneously: VMAF (\texttt{vmaf\_v0.6.1}) \cite{netflix_vmaf}, VMAF-NEG (\texttt{vmaf\_v0.6.1neg}) \cite{netflix_vmafneg}, PSNR, and MS-SSIM.

\paragraph{BD-rate.} Piecewise Cubic Hermite Interpolating Polynomial (PCHIP) fit through four RD points, Bj{\o}ntegaard delta \cite{bjontegaard2001bdrate} integrated with \texttt{scipy.integrate.quad}, arithmetic mean across sequences. Negative BD-rate means bitrate saved at matched quality; positive BD-quality means quality gained at matched bitrate.

\paragraph{Two legs per sequence.} (1)~\textbf{Baseline}: \texttt{original.yuv} $\to$ \texttt{libx264} $\to$ \texttt{mp4}. (2)~\textbf{Kelvin}: \texttt{original.yuv} $\to$ Kelvin (Mode~C, preprocessing only) $\to$ \texttt{preprocessed.yuv} $\to$ \texttt{libx264} $\to$ \texttt{mp4}. Both are decoded back to YUV and scored against the \emph{original} raw YUV. The only variable is the preprocessor.

\subsection{UVG, 1080p}
\label{sec:uvg}

Seven sequences, $1920{\times}1080$ at 120 fps \cite{mercat2020uvg}. The headline number is BD-VMAF mean $\boldsymbol{-27.62\%}$, std 7.64, range $[-39.83\%, -16.40\%]$, with every sequence negative.

\begin{table}[h]
\centering
\caption{UVG per-sequence Bj{\o}ntegaard-delta rates against baseline \texttt{libx264} \emph{medium}. Negative = bitrate saved at matched quality. Source: \texttt{kelvin-benchmark/results/uvg\_summary.csv}.}
\label{tab:uvg}
\small
\begin{tabular}{lrrrr}
\toprule
Sequence & BD-VMAF & BD-VMAF-NEG & BD-PSNR-Y & BD-MS-SSIM \\
\midrule
Beauty         & $-39.83\%$ & $-9.84\%$ & $+39.20\%$ & $+20.00\%$ \\
Bosphorus      & $-27.20\%$ & $-6.52\%$ & $+15.58\%$ & $+5.15\%$  \\
HoneyBee       & $-33.67\%$ & $+2.69\%$ & $+35.29\%$ & $+22.59\%$ \\
Jockey         & $-20.83\%$ & $-5.69\%$ & $+20.38\%$ & $+10.75\%$ \\
ReadySteadyGo  & $-16.40\%$ & $-4.57\%$ & $+10.61\%$ & $+3.65\%$  \\
ShakeNDry      & $-28.14\%$ & $-4.62\%$ & $+17.78\%$ & $+6.63\%$  \\
YachtRide      & $-27.31\%$ & $-7.69\%$ & $+10.80\%$ & $+2.11\%$  \\
\midrule
\textbf{mean (n=7)} & $\boldsymbol{-27.62\%}$ & $\boldsymbol{-5.18\%}$ & $\boldsymbol{+21.38\%}$ & $\boldsymbol{+10.13\%}$ \\
\bottomrule
\end{tabular}
\end{table}

\paragraph{Reading the table.} VMAF is the perceptual quality metric the streaming industry uses; the headline savings map onto VMAF. VMAF-NEG is Netflix's gaming-resistant variant, designed to penalize pre-processing that artificially inflates perceptual scores. PSNR-Y is pixel-wise luma fidelity; the $+21.38\%$ cost is the by-design trade-off, since Kelvin spends bits on perceptual structure and pays back with bits on pixel-wise fidelity. MS-SSIM is multiscale structure and tracks PSNR in direction.

\textbf{7 of 7 wins on BD-VMAF; 6 of 7 wins on BD-VMAF-NEG; 7 of 7 wins on BD-PSNR-Y.} The gain ranges from $-16.4\%$ on the highest-motion clip (ReadySteadyGo) to $-39.8\%$ on Beauty. The single positive BD-VMAF-NEG (HoneyBee, $+2.69\%$) is small and surrounded by very large gains on the standard model and PSNR-Y; it does not look like model gaming. We keep HoneyBee in the corpus rather than excluding it; the divergence is itself the most informative thing the table publishes.

\paragraph{The Beauty operating point.} The cleanest equivalent-quality read on UVG comes from Beauty. The baseline RD curve hits VMAF 92.48 at $\mathrm{QP}=22$ at 101{,}930 kbps. Kelvin+\texttt{libx264} reaches VMAF 91.76 at $\mathrm{QP}=27$ at 19{,}612 kbps: $80.8\%$ bitrate reduction at $0.7$ VMAF below the matched-quality point. Integrated across the four-point RD curve, the BD-rate is the $-39.83\%$ in Table~\ref{tab:uvg}.

\begin{figure}[h]
\centering
\includegraphics[width=0.95\linewidth]{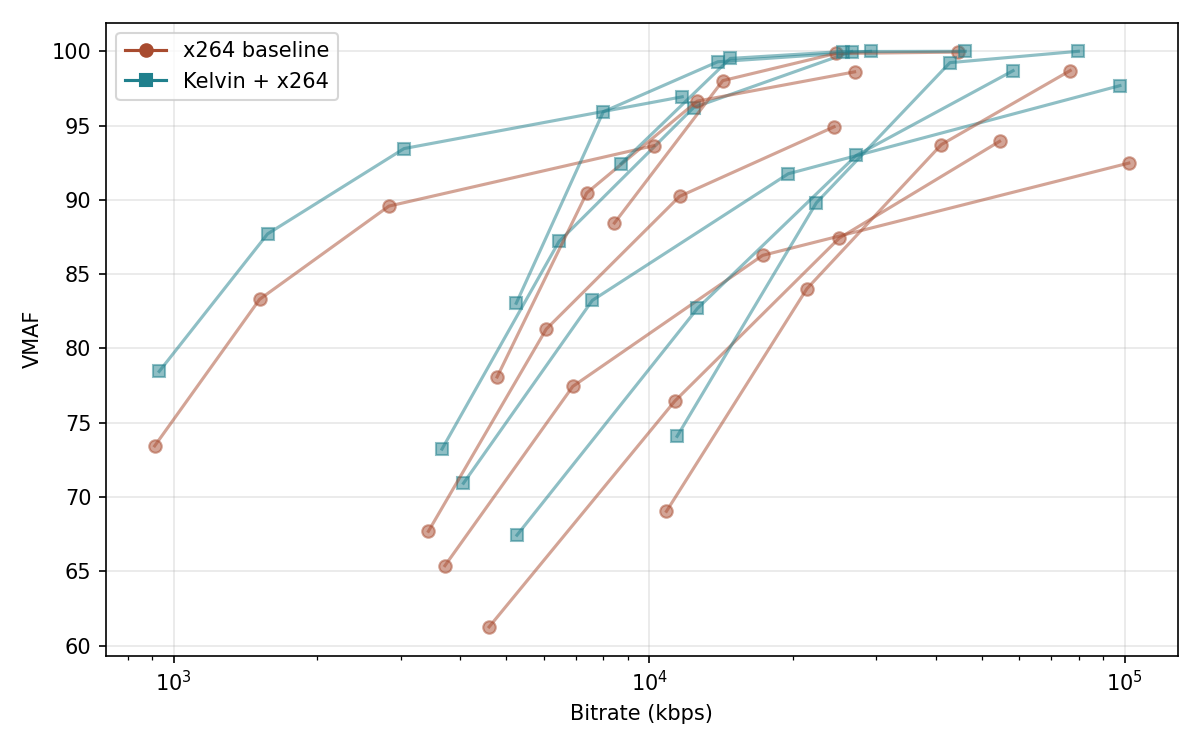}
\caption{UVG rate-distortion curves under VMAF, all 7 sequences. Baseline \texttt{libx264} \emph{medium} (open markers) vs.\ Kelvin~+~\texttt{libx264} (filled markers). Kelvin curves sit above-left of baseline on every sequence (lower bitrate at the same VMAF, or higher VMAF at the same bitrate).}
\label{fig:uvg_combined_vmaf}
\end{figure}

\begin{figure}[h]
\centering
\includegraphics[width=0.95\linewidth]{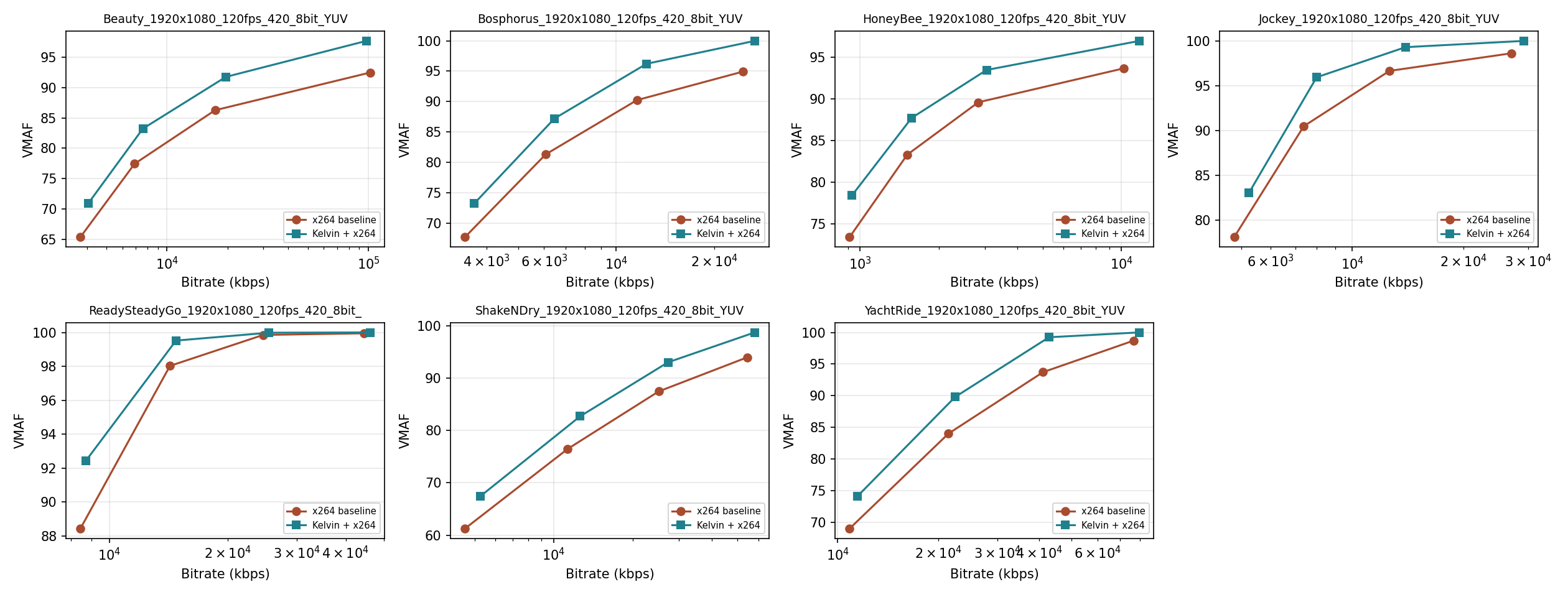}
\caption{UVG per-sequence VMAF RD curves. Every panel shows Kelvin sitting above-left of baseline. The vertical extent of each panel is the four-QP grid $\{22, 27, 32, 37\}$.}
\label{fig:uvg_per_seq_vmaf}
\end{figure}

\subsection{MCL-JCV generalization}
\label{sec:mcljcv}

To estimate how the UVG number generalizes to unseen content, we ran the same harness on 30 sequences of the public MCL-JCV dataset \cite{wang2016mcljcv}. Same checkpoint, same codec, same QPs, same metrics. The result is interesting in both directions, and the only honest publication is in two cuts.

\begin{table}[h]
\centering
\caption{MCL-JCV aggregates. Negative = bitrate saved at matched quality. Source: \texttt{kelvin-benchmark/results/mcljcv\_summary.csv}.}
\label{tab:mcljcv}
\small
\begin{tabular}{lrrrr}
\toprule
Slice & n & BD-VMAF & BD-VMAF-NEG & BD-PSNR-Y \\
\midrule
All clips                          & 30 & $-16.83\%$ & $+4.42\%$   & $+24.86\%$ \\
Median (all)                       & 30 & $-25.39\%$ & $-5.64\%$   & $+15.60\%$ \\
Excl.\ \texttt{SRC09}, \texttt{SRC13} (regressions) & 28 & $\boldsymbol{-27.70\%}$ & $\boldsymbol{-5.37\%}$ & $+20.56\%$ \\
Excl.\ \texttt{SRC09}, \texttt{SRC13}, \texttt{SRC29} & 27 & $-26.65\%$ & $-5.08\%$ & $+18.38\%$ \\
\bottomrule
\end{tabular}
\end{table}

\paragraph{Reading the table.} Kelvin wins on 28 of 30 unseen MCL-JCV clips on BD-VMAF. The median ($-25.39\%$) is the typical clip; the $n=30$ mean ($-16.83\%$) is dragged down by two outliers. Drop those two and BD-VMAF-NEG flips from $+4.42\%$ to $-5.37\%$; the gaming-resistant metric agrees with the perceptual win on the non-pathological 28 of 30. The sign flip is the most informative line in the table.

\paragraph{The two datasets line up.} UVG mean is $-27.62\%$ BD-VMAF (Section~\ref{sec:uvg}). MCL-JCV $n=28$ mean is $-27.70\%$. That is a $\sim 0.1$ percentage point gap across two independent test sets, one curated, one unseen, on the same checkpoint with the same harness. This is Kelvin v1.0's expected H.264 performance on natural video at iso-VMAF, away from the named failure modes.

\begin{figure}[h]
\centering
\includegraphics[width=0.85\linewidth]{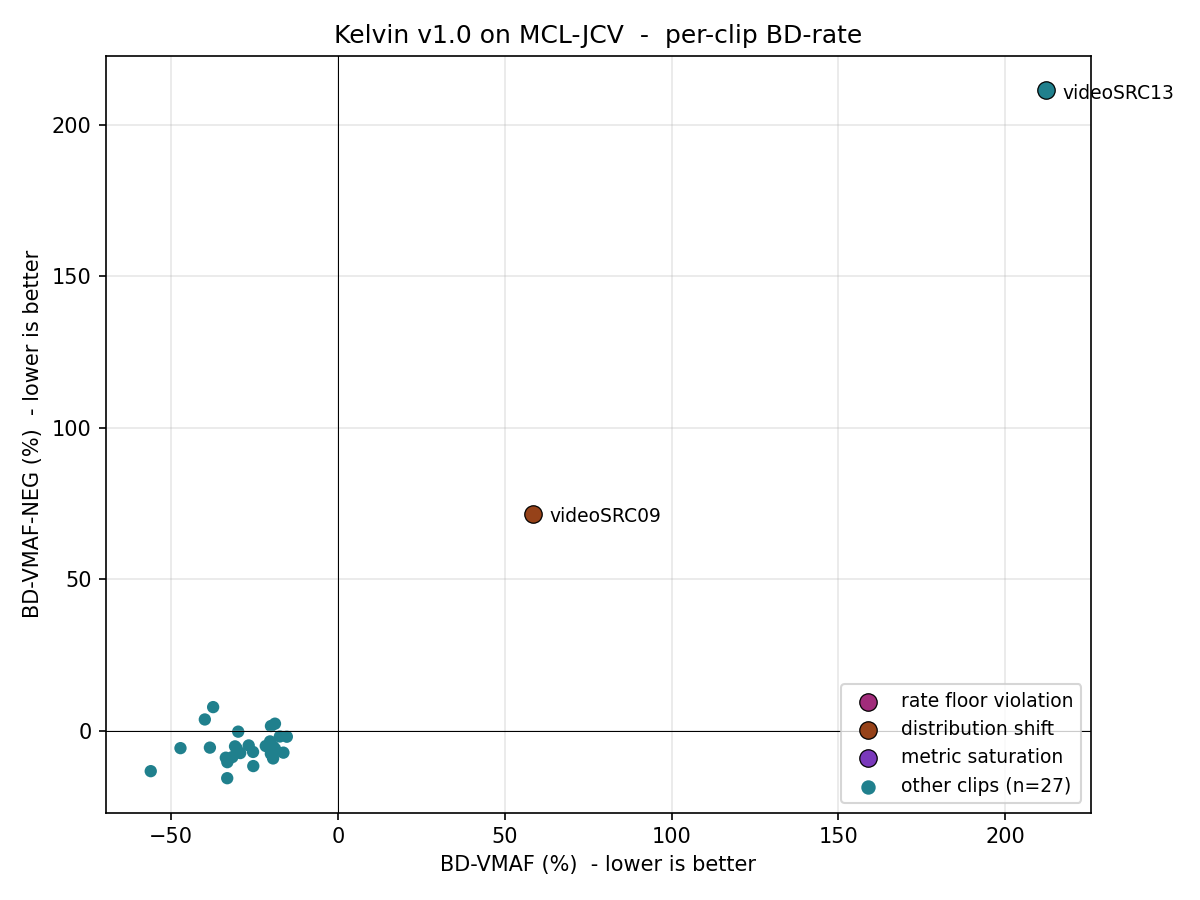}
\caption{MCL-JCV BD-VMAF vs.\ BD-VMAF-NEG scatter, all 30 clips. The lower-left quadrant contains clips where Kelvin saves bitrate under both the standard and the gaming-resistant perceptual model; the two upper-right points are the named regressions (Section~\ref{sec:failures}).}
\label{fig:mcljcv_scatter}
\end{figure}

\subsection{Failure-mode taxonomy}
\label{sec:failures}

Each named outlier maps to a distinct failure class with a corresponding v1.1 fix. Promoting that mapping to a taxonomy makes it useful as a diagnostic framework for the next dataset, not just a description of this one.

\paragraph{\texttt{videoSRC13}, rate-floor violation.} Landscape dominated by smooth, near-uniform sky. \textbf{BD-VMAF $+212.23\%$, BD-VMAF-NEG $+211.35\%$.} Baseline \texttt{libx264} was encoding the smooth sky for near-zero bits per macroblock; any Kelvin modification in a region the encoder was already encoding for free is bits added with no quality return. v1.1 fix: an explicit smooth-content gate.

\paragraph{\texttt{videoSRC09}, distribution shift.} A pedestrian mall with a saturated-chroma tulip foreground; near-monochrome chroma planes that sit outside the Vimeo-90K training distribution. \textbf{BD-VMAF $+58.48\%$, BD-VMAF-NEG $+71.61\%$.} v1.1 fix: a broader chroma corpus.

\paragraph{\texttt{videoSRC29} (historical, now retained).} Low-light cinematic interior at 24 fps. On an earlier checkpoint we excluded this clip as ``metric saturation'' because the baseline VMAF was high enough that the integration window pressed against the 100-cap. Under the v12 production checkpoint reported here it reads BD-VMAF $-56.09\%$, BD-VMAF-NEG $-13.33\%$, and BD-PSNR $+79.40\%$; the underlying gain is real and large, so we retain it in the canonical $n=28$ cut. The $n=27$ slice is preserved for compatibility with prior reporting.

Naming the failure modes is more useful to a buyer than hiding them. The customer who shoots stock landscape footage learns from \texttt{SRC13} in 30 seconds whether Kelvin v1.0 fits their library.

\subsection{Per-sequence and additional plots}

\begin{figure}[h]
\centering
\begin{minipage}[t]{0.48\linewidth}
\includegraphics[width=\linewidth]{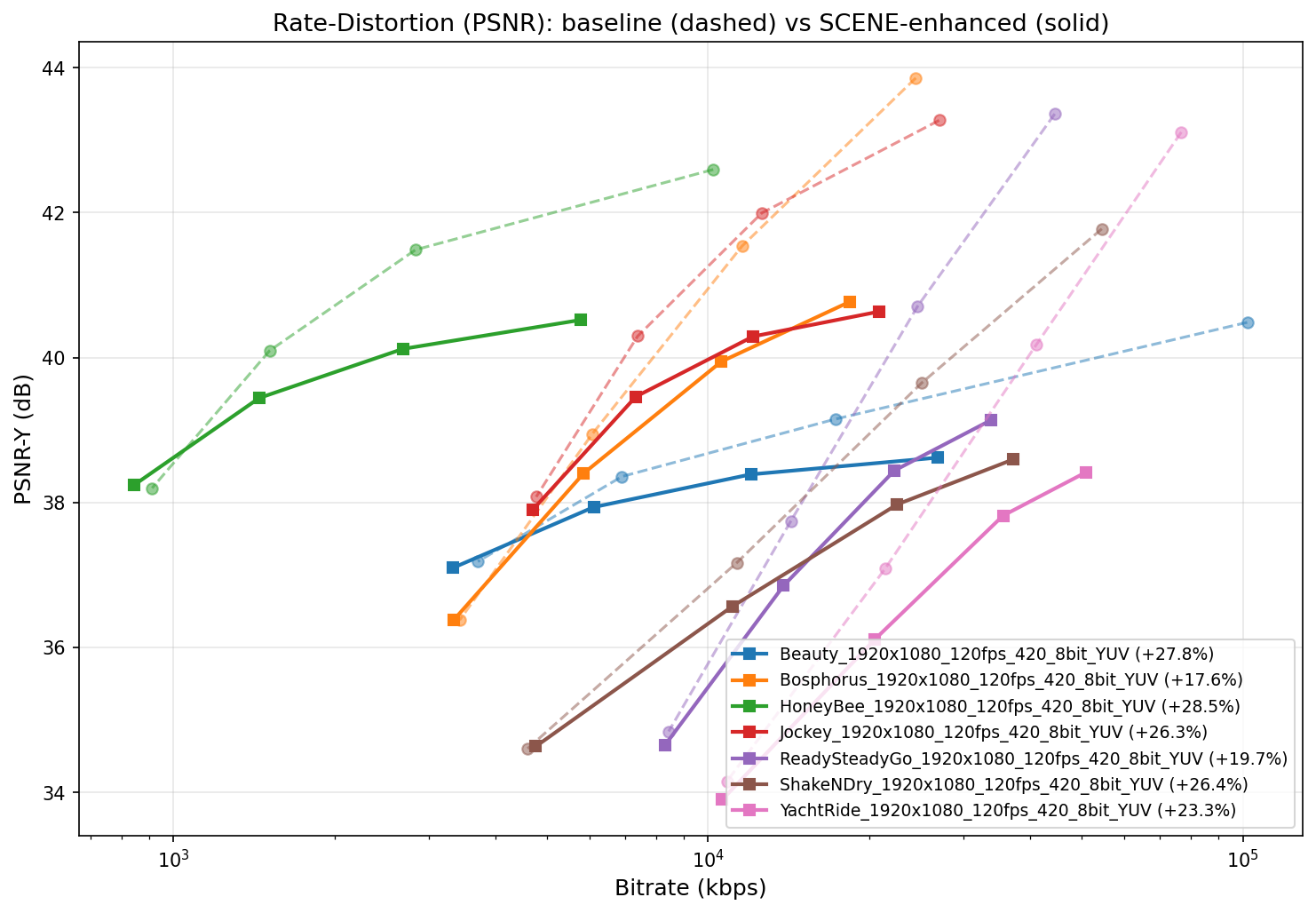}
\subcaption{Combined PSNR.}
\end{minipage}\hfill
\begin{minipage}[t]{0.48\linewidth}
\includegraphics[width=\linewidth]{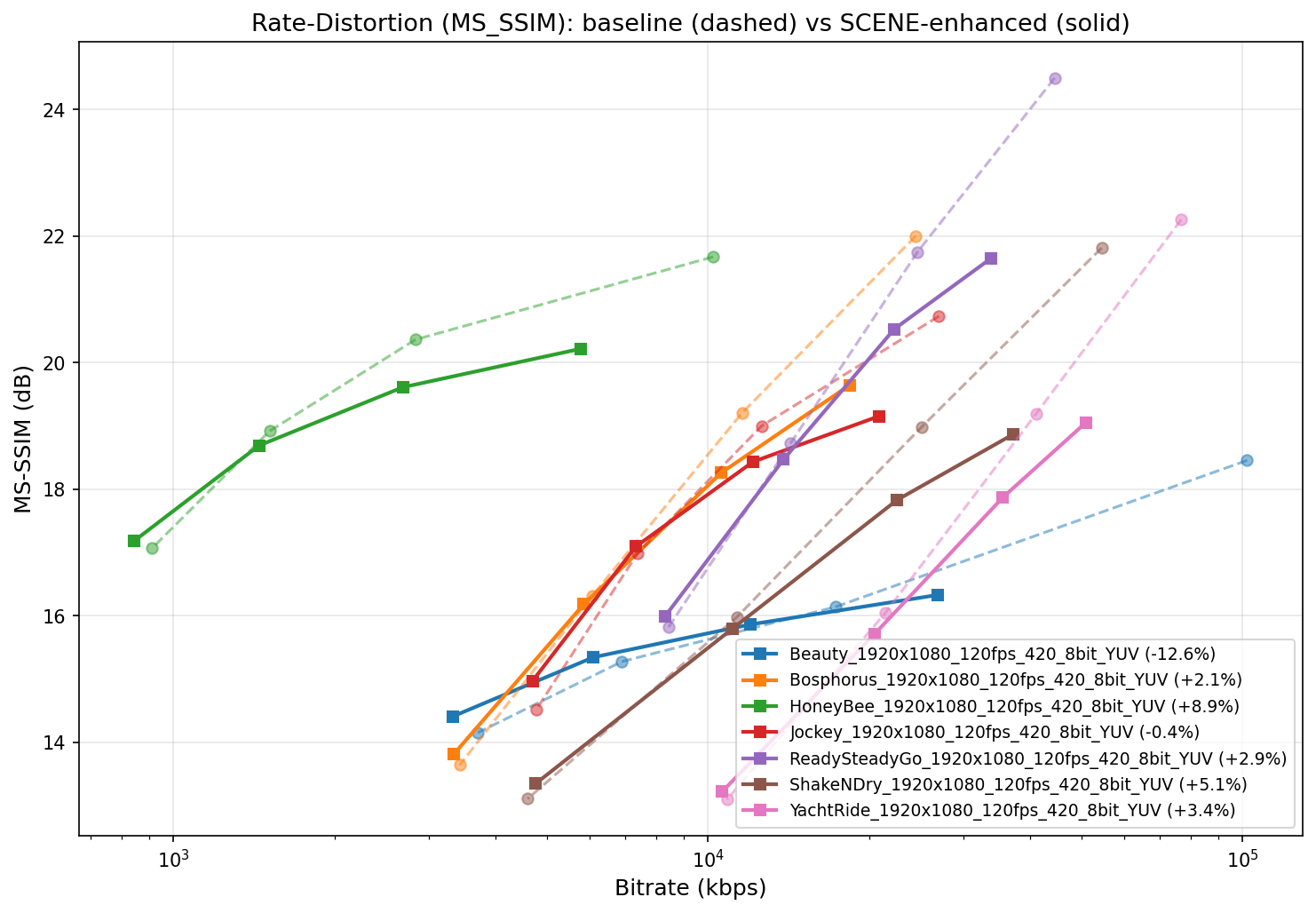}
\subcaption{Combined MS-SSIM.}
\end{minipage}
\caption{UVG combined RD curves under PSNR-Y and MS-SSIM. Kelvin trades pixel-wise fidelity ($+21.38\%$ BD-PSNR mean) and structural fidelity ($+10.13\%$ BD-MS-SSIM mean) for perceptual quality. Workloads requiring pixel-exact preservation should use plain \texttt{libx264}.}
\label{fig:uvg_combined_other}
\end{figure}

\begin{figure}[h]
\centering
\includegraphics[width=\linewidth]{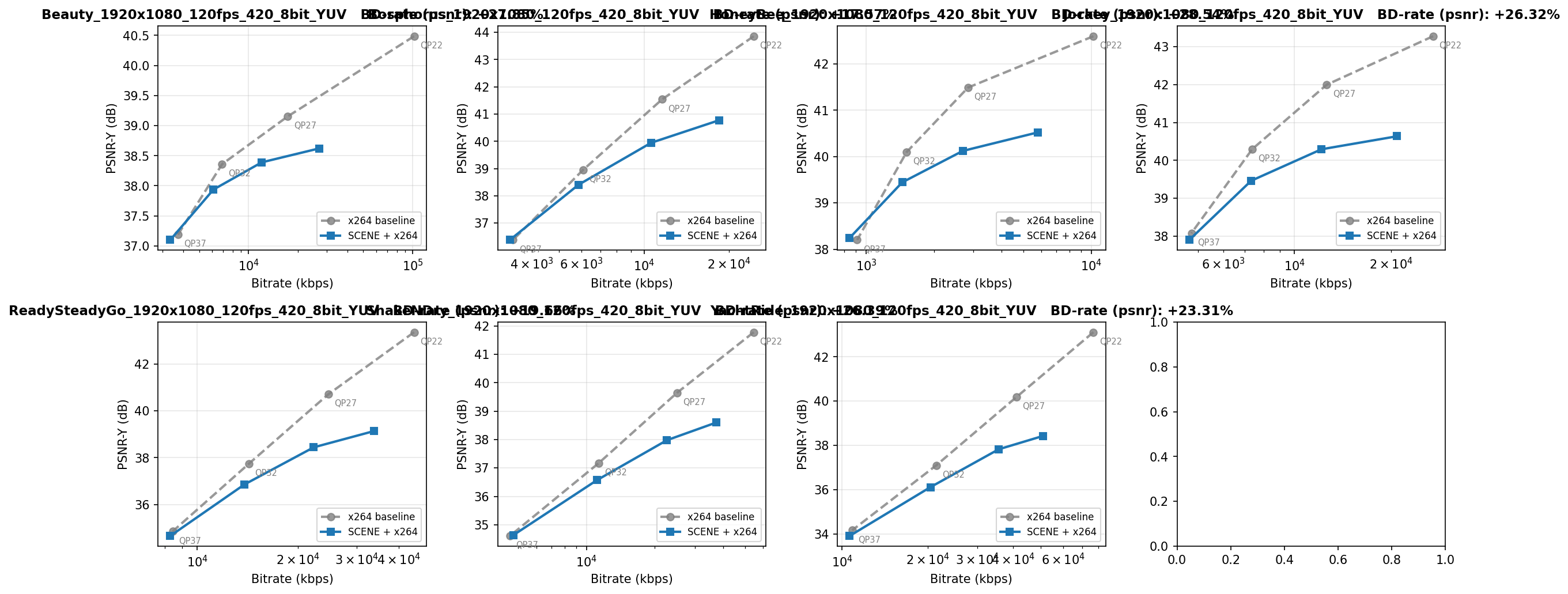}
\caption{UVG per-sequence RD curves under PSNR-Y.}
\label{fig:uvg_per_seq_psnr}
\end{figure}

\begin{figure}[h]
\centering
\includegraphics[width=\linewidth]{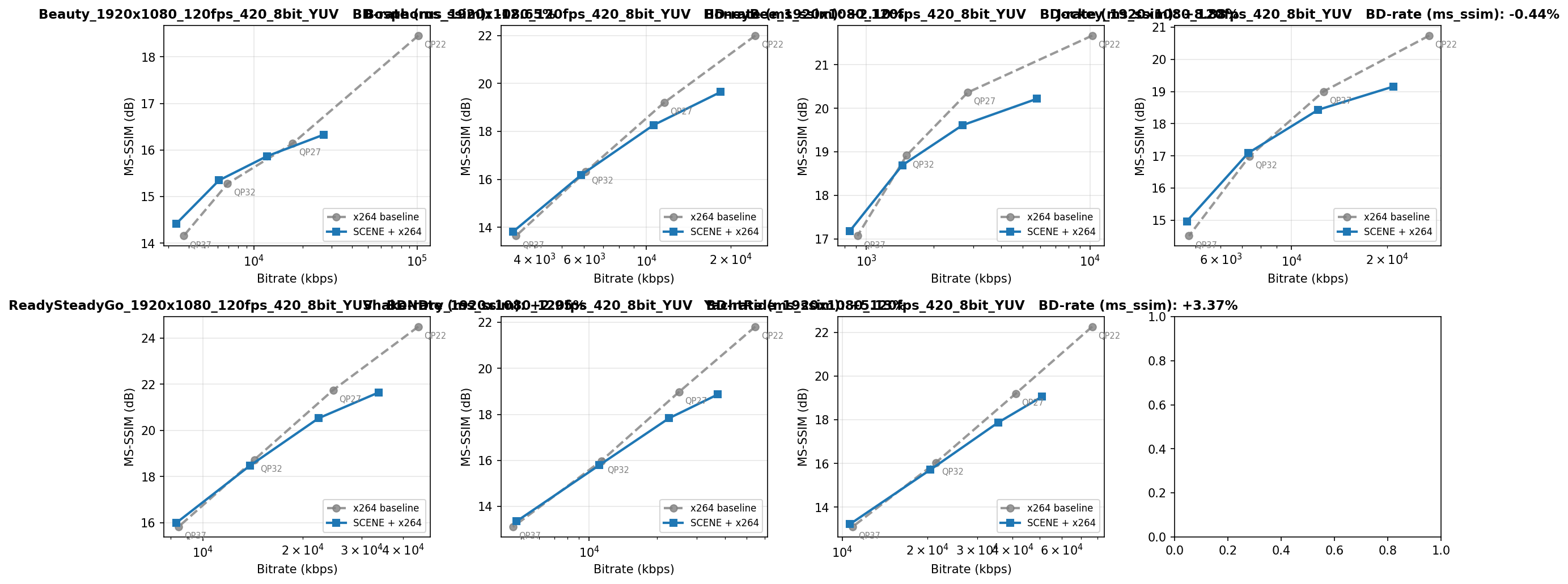}
\caption{UVG per-sequence RD curves under MS-SSIM.}
\label{fig:uvg_per_seq_msssim}
\end{figure}

\section{Baseline sanity panel}
\label{sec:baselines}

A $-27.62\%$ BD-rate on VMAF invites the obvious question: compared to what. ``Plain \texttt{libx264} \emph{medium}'' is the right denominator for a pre-encoder claim, but not the only reference. We therefore ran the same harness against five additional baselines on the same UVG corpus: two classical prefilters (\texttt{hqdn3d}, \texttt{unsharp}), two perceptual-tuning flags (\texttt{-tune psnr}, \texttt{-tune ssim}), and the next-codec-up (x265 \emph{medium}). Every baseline runs at vanilla / default parameters.

\begin{table}[h]
\centering
\caption{Baseline sanity panel on UVG ($n=7$). All numbers are BD-rate against plain \texttt{libx264} \emph{medium} except the x265 row which is x265 \emph{medium} vs.\ \texttt{libx264} \emph{medium}. Negative = bitrate saved. Source: companion sweep to \cite{kelvinbench2026,grazianomedium2026}.}
\label{tab:baselines}
\small
\begin{tabular}{lrrrr}
\toprule
Variant & BD-VMAF & BD-VMAF-NEG & BD-PSNR-Y & BD-MS-SSIM \\
\midrule
\texttt{libx264 + hqdn3d}    & $+24.94\%$ & $+25.10\%$ & $+25.30\%$ & $+25.20\%$ \\
\texttt{libx264 + unsharp}   & $-21.67\%$ & $\boldsymbol{+20.27\%}$ & $\boldsymbol{+139.02\%}$ & $+45.10\%$ \\
\texttt{libx264 -tune psnr}  & $-9.38\%$  & $-10.96\%$ & $-17.01\%$ & $-15.94\%$ \\
\texttt{libx264 -tune ssim}  & $-9.38\%$  & $-10.96\%$ & $-17.01\%$ & $-15.94\%$ \\
\textbf{Kelvin + libx264}    & $\boldsymbol{-27.62\%}$ & $\boldsymbol{-5.18\%}$ & $+21.38\%$ & $+10.13\%$ \\
\midrule
x265 \emph{medium} (next codec) & $-32.50\%$ & $-32.62\%$ & $-39.36\%$ & $-40.72\%$ \\
\bottomrule
\end{tabular}
\end{table}

\paragraph{\texttt{hqdn3d}.} Default denoising is uniformly worse on every metric. Denoising removes structure the encoder was already representing efficiently. Kelvin is not a denoiser.

\paragraph{\texttt{unsharp}, the textbook VMAF-gaming case.} BD-VMAF $-21.67\%$ (superficially close to Kelvin's $-27.62\%$) but \textbf{BD-VMAF-NEG $+20.27\%$ on 7 of 7 sequences} and BD-PSNR $+139.02\%$. Unsharp injects high-frequency content that VMAF rewards as ``sharper''; VMAF-NEG, built specifically to discount such enhancements, rejects the gain. The categorical distinction is the agreement direction across metrics, not the magnitude of any single one. Kelvin: VMAF-NEG $-5.18\%$ on 6 of 7 sequences; BD-PSNR $+21.38\%$, an order of magnitude smaller cost than unsharp. The Kelvin training objective penalizes bitrate inflation, and the $\delta_{\max}$ cap prevents the network from injecting visible energy; \texttt{unsharp} has neither constraint.

\paragraph{\texttt{-tune psnr / -tune ssim}.} Modest, uniformly negative BD-rates. Under \texttt{libx264}'s CQP rate control these two tunes produce byte-identical bitstreams (verified by MD5); the only flag that differs is \texttt{--aq-mode 1} vs.\ \texttt{--aq-mode 2}, and adaptive quantization has no effect under fixed-QP encoding. The duplication is kept for transparency.

\paragraph{x265 \emph{medium}.} This is the most important row in the panel for honest positioning. x265 beats Kelvin on every metric, and it does so without the metric-divergence pattern: BD-VMAF-NEG ($-32.62\%$) tracks BD-VMAF ($-32.50\%$) almost exactly. No perceptual gaming, just a genuinely better codec.

\paragraph{Positioning.} If your pipeline can move to HEVC, you should. Kelvin's value is not ``Kelvin beats HEVC at H.264 BD-rate''; it does not. Kelvin's value is the workloads where moving to HEVC is not a free choice: legacy decoder reach (universal H.264 decode is a property no other codec has), contractual H.264 mandates, CDN constraints, and content libraries already encoded in H.264. That market still represents the majority of internet video traffic \cite{bitmovin2025}. Within that constraint, the relevant comparison is Kelvin ($-27.62\%$ BD-VMAF) vs.\ the next-best H.264 alternative we measured, \texttt{-tune psnr} ($-9.38\%$ BD-VMAF): a roughly $3\times$ larger saving on the same codec. That is the comparison Kelvin wins, and it is the comparison that defines its product surface.

\section{Limitations and roadmap}
\label{sec:limitations}

A short, declarative list of what v1.0 does not do, each mapped to a v1.1 roadmap item.

\begin{itemize}[leftmargin=*]
    \item \textbf{Frame-by-frame, no temporal modeling.} The single largest architectural gap. Temporal flicker is mitigated by the global residual skip and the tight $\delta$-cap but not eliminated. Tracked under inter-frame awareness.
    \item \textbf{H.264 only.} The proxy is calibrated to \texttt{libx264} preset \emph{medium}, single-threaded. HEVC, AV1, and other x264 presets each require a separate proxy retrain; the model itself is otherwise codec-agnostic.
    \item \textbf{8-bit, 1080p, BT.709/BT.601.} 10-bit and HDR are roadmap items.
    \item \textbf{VOD-first, with early WebRTC evidence.} Single-L40S throughput is roughly 35 fps at 1080p ($\sim 28$ ms per frame); processing one hour of 1080p takes $\sim 103$ GPU-seconds. As a first step beyond the VOD scope, Kelvin has been wired into a WebRTC test harness running the same feed through two parallel \texttt{aiortc} peer connections (passthrough vs.\ Kelvin) with per-second VMAF measured server-side. End-to-end per-frame latency through the harness is $27.6$ ms p50 / $29.0$ ms p95 at 720p (fp16, \texttt{cudnn.benchmark=True}, RTX-class GPU). An iso-quality controller drives Kelvin's encoder target downward until smoothed VMAF matches the baseline tile; on stable indoor scenes at a 500 kbps baseline it settles at $\sim 390$ kbps, i.e.\ $\sim 22\%$ bandwidth saving at matched quality. The harness exercises the streaming path but does not yet cover keyframe alignment with upstream packagers, scene-cut handling, or SFU integration.
    \item \textbf{SigLIP~2 dependency.} $\sim 1.8$ GB on disk and $\sim 400$M inference parameters. A SigLIP-free distillation and per-shot embedding caching will materially reduce footprint without changing the BD-rate claim.
    \item \textbf{Ablation table not yet published.} The $\delta$-cap, multiplicative-only FiLM, and ReLU-vs-GELU design decisions are asserted with internal evidence but not yet published as a side-by-side BD-VMAF ablation table on the production UVG harness. Each variant requires a fresh Kelvin retrain with identical hyperparameters and an identical UVG sweep; we prefer to publish the table once validated rather than estimate it now.
\end{itemize}

\section{Conclusion}

Kelvin v1.0 demonstrates that standards-compliant neural pre-processing can deliver substantial bitrate savings in production H.264 pipelines while preserving deployment simplicity. On UVG the mean BD-VMAF is $-27.62\%$ (7 of 7 wins, std 7.64); on MCL-JCV the same checkpoint wins on 28 of 30 unseen clips at $-27.70\%$ BD-VMAF in the named-failure-excluded cut. The two datasets line up within $\sim 0.1$ percentage points. The single $\pm 1/255$ residual constraint and the codec-faithful distortion proxy together produce gains that survive the gaming-resistant VMAF-NEG model in the majority of cases, separating Kelvin from classical sharpening filters whose VMAF-NEG signature is uniformly positive. Against x265 \emph{medium} on the same corpus Kelvin loses on every metric, which is the honest result and shapes the product surface: Kelvin is the best H.264 option in our panel for the workloads where remaining on H.264 is a constraint. The full benchmark harness, configs, CSVs, and plots are released at \texttt{github.com/marcoeg/kelvin-benchmark} \cite{kelvinbench2026}, and a companion technical writeup is at \cite{grazianomedium2026}. The model ships in the EncodeIQ cloud service (\url{https://www.encodeiq.ai}).

\paragraph{Reproducibility.} Tier-1 reproduction (anyone, $\sim 30$ minutes): take the published encoded \texttt{.mp4} bitstreams, the original UVG/MCL-JCV YUVs, the \texttt{libvmaf} invocation in \texttt{configs/libvmaf.json}, and \texttt{scripts/bjontegaard.py}. The four CSVs in \texttt{results/} are pre-computed for convenience; running \texttt{python scripts/bjontegaard.py results/uvg\_rd\_per\_qp\_vmaf.csv} reproduces the UVG mean of $-27.62\%$ to within rounding.

\paragraph{Acknowledgments.} The architectural family draws on SCENE \cite{lin2026scene}; the conditioning backbone is SigLIP~2 \cite{tschannen2025siglip2}. We thank the maintainers of UVG \cite{mercat2020uvg} and MCL-JCV \cite{wang2016mcljcv} for releasing the datasets used in this evaluation, and the Netflix VMAF team \cite{netflix_vmaf,netflix_vmafneg} for the perceptual model that this report leans on throughout.

\bibliographystyle{plain}
\bibliography{refs}

\end{document}